\documentclass[twocolumn,prd,showpacs,superscriptaddress,groupedaddress,preprintnumbers,nofootinbib]{revtex4-1}
\pdfoutput=1
\usepackage{graphicx,amsfonts,amsmath,amssymb,epsfig,color, mathrsfs,latexsym,url,wasysym,float}
\newcommand{\be}{\begin{equation}}
\newcommand{\ee}{\end{equation}}
\newcommand{\nn}{\nonumber}
\newcommand{\ba}{\begin{eqnarray}}
\newcommand{\ea}{\end{eqnarray}}
\newcommand{\mpl}{m_{\rm Pl}}
\begin{document}

\title{dRGT Theory of Massive Gravity from Spontaneous Symmetry Breaking}

\author{Mahdi Torabian$^*$}
\affiliation{Department of Physics, Sharif University of Technology, Azadi Ave, 11155-9161, Tehran, Iran}

\begin{abstract}
In this note we propose a topological action for a Poincare times diffeomorphism invariant gauge theory. 
We show that there is Higgs phase where the gauge symmetry is spontaneous broken to a diagonal Lorentz subgroup and gives the Einstein-Hilbert action plus the dRGT potential terms. In this vacuum, there are five (three from Goldstone modes) propagating degrees of freedom which form polarizations of a massive spin 2 particle, an extra healthy heavy scalar (Higgs) mode and no Boulware-Deser ghost mode. We further show that the action can be derived in a limit from a topological de Sitter invariant gauge theory in 4 dimensions. 
\end{abstract}

\pacs{}
\maketitle

\subsection*{Introduction} 
The construction of a local and Lorentz invariant field theory for interacting massive spin 2 particles has long been an interesting academic question. It is cosmologically interesting too which modifies the behavior of gravity at long distances and offer a dynamical explanation for the late time accelerated expansion of the Universe. Applying the effective field theory view on a theory of massive gravity \cite{ArkaniHamed:2002sp}, the authors of \cite{deRham:2010ik,deRham:2010kj} constructed a potential contribution to the Einstein-Hilbert action which gives graviton a mass and modifies the dynamics of General Relativity in the IR limit. This fully non-linear theory ({\it a.k.a.} dRGT theory) propagates five degrees of freedom for each polarization of massive spin 2 particle and avoids the Boulware-Deser ghost excitation \cite{Boulware:1973my} by carefully tuning interactions \cite{deRham:2010ik,deRham:2010kj,Hassan:2011hr,Hassan:2011ea}.  

The existence of a fiducial fixed Minkowski metric in the dRGT theory, breaks diffeomorphism invariance of the Einstein-Hilbert action and makes longitudinal/Goldstone modes dynamical. However being derivatively coupled, the Goldstone interactions grow with energy and get stronger that eventually break perturbation theory at some scale. This cut-off scale of the effective field theory is simply determined by computing the tree level scattering amplitudes and seeing when it hits the unitarity bound. Around Minkowski spacetime, the cut off is $\Lambda_3=(m^2\mpl)^{1/3}$ which is unacceptably low about $10^{-13}$ eV for phenomenologically motivated value for the graviton mass $m\sim H_0\sim 10^{-33}$ eV. The common lore is that this low energy effective field theory must be extended at higher energies to its UV completion with possibly new degrees of freedom. 

On the one hand, one would wonder if the Lorentz invariant massive gravity could have a Lorentz invariant UV completion. Recently, a nice study showed that there is a region in the parameter space of the dRGT theory where it is indeed possible \cite{Cheung:2016yqr}. In fact by computing four particles forward scattering amplitudes it is shown that analyticity and unitarity conditions are satisfied within this window. 

On the other hand, in the case of massive non-Abelian gauge theories, the Higgs mechanism is responsible for dynamical symmetry breaking and giving mass to vector fields in a weakly coupled regime. Furthermore, it UV completes the theory by introducing a heavy extra Higgs mode. Analogously, one would wonder if there is a microscopic theory which leads to the dRGT theory through spontaneous gauge symmetry breaking and dynamically generates mass. In order to approach this idea, one first needs to formulate gravity as a gauge theory and then construct a mechanism for dynamical symmetry breaking. Gauge theory for gravity has a long history and many attempts were made to construct a model (see \cite{Utiyama:1956sy,Kibble:1961ba,Sciama:1964wt,MacDowell:1977jt,Chamseddine:1976bf,Hayashi:1979wj,Ivanov:1981wn,Ivanenko:1984vf,Hehl:1994ue}). However, a satisfactory symmetry breaking mechanism to produce massive graviton (\`{a} l{a} dRGT) in still lacking (see \cite{Percacci:1984ai,Percacci:1990wy,Kirsch:2005st,tHooft:2007rwo,Chamseddine:2010ub,Alberte:2010it,Alberte:2010qb,Dubovsky:2004sg} for previous attempts). 

In this note, we propose a dynamical mechanism through which the spin 2 particles receive mass. We start from a topological action which is invariant under local Poincare transformations times diffeomorphisms. The Poincare group is especially motivated as it is the symmetry of Minkowski spacetime that well approximates the visible Universe. The dynamical fields are gauge connections plus spacetime scalars in vector and singlet representations of the Lorentz group. We show that there exist a Higgs phase where scalars (and derivative of them) receive non-zero vacuum expectations values and thereby breaks the gauge symmetry to a diagonal Lorentz subgroup. In this phase we find the Einstein-Hilbert action plus the dRGT potential terms. Around this vacuum there are 5 degrees of freedom of a massive spin 2 particle, 3 of which are Goldstone modes from the extra vector, and a massive healthy heavy Higgs mode. The Boulware-Deser ghost is removed in the special form of the potential. The Higgs excitation can help to improve the UV behavior of longitudinal polarizations of the massive tensor mode. Finally, we show that the action can be derived from a simpler looking topological de Sitter invariant action. 

\paragraph*{The dRGT Potential Terms}
Before we move on to introduce the action we briefly review the dRGT theory \cite{deRham:2010ik,deRham:2010kj}. Through careful tuning of all possible interactions, a fully non-linear theory for massive graviton has been proposed   which propagates 5 healthy degrees of freedom  \cite{Hassan:2011hr,Hassan:2011ea}. In terms of vierbein 1-forms $e^a=e^a_\mu {\rm d}x^\mu$ and a unit 1-form ${\bf 1}^a=\delta^a_\mu{\rm d}x^\mu$ the potential can be written into an interesting form as follows \cite{Hinterbichler:2012cn}
\ba\label{dRGT-potential} \frac{-V^{\rm dRGT}}{m^2\mpl^2} &=& \frac{b_1}{6}\epsilon^{abcd}e^a\wedge e^b\wedge e^c\wedge {\bf 1}^d\cr
&+&\frac{b_2}{4}\epsilon^{abcd}e^a\wedge e^b\wedge {\bf 1}^c\wedge {\bf 1}^d\cr
&+&\frac{b_3}{6}\epsilon^{abcd}e^a\wedge {\bf 1}^b\wedge {\bf 1}^c\wedge {\bf 1}^d,
\ea 
where $m$ is the graviton mass given that $b_1+2b_2+b_3 = 1$. This particular form of the potential gives a constraint that projects out the ghost-like degree of freedom. The cosmological constant term can also be added to the above potential as $(b_0/4!)\epsilon_{abcd} e^a\wedge e^b\wedge e^c\wedge e^d$.
\vspace*{-1mm}
\subsection*{The Action}
We start with a topological gauge theory which is invariant under local Poincare transformations and spacetime diffeomorphisms $ISO(1,3)\times${\it Diff}. Needless to say with no metric in the action, diffeomorphism invariance is essentially topological invariance. The Poincare group is an internal group and diffeomorphisms act over 4 dimensional base (spacetime) manifold. No metric structure on the base manifold is assumed a priori, however, it can be defined a posteriori. The action reads as
\ba\label{Poincare-action} S = l\epsilon_{abcd}\int_{\cal M}  &c_1&\varphi R^{ab}\wedge R^{cd} \nn\\ 
+&c_2& \varphi R^{ab}\wedge({\rm D}\phi^c+\varphi e^c)\wedge ({\rm D}\phi^d+\varphi e^d) \nn\\
+&c_3& \varphi({\rm D}\phi^a+\varphi e^a)\wedge({\rm D}\phi^b+\varphi e^b)\nn\\&&\qquad\qquad \wedge ({\rm D}\phi^c+\varphi e^c)\wedge({\rm D}\phi^d+\varphi e^d)\nn\\
-2 &c_2&R^{ab} \wedge ({\rm D}\phi^c+\varphi e^c)\wedge {\rm d}\varphi  \phi^d \nn\\
-4 &c_3&({\rm D}\phi^b+\varphi e^b)\wedge({\rm D}\phi^c+\varphi e^c\big) \nn\\
&&\qquad\qquad \wedge({\rm D}\phi^c+\varphi e^c)\wedge {\rm d}\varphi  \phi^d\nn\\
-l\int_{\cal M}&g&(\varphi^2-v^2)^2\Lambda_4
.\ea
The dynamical fields are Lorentz $\omega^{ab}$ and translation  $e^a$ 1-form connections (gauge fields for respective transformations) as well as Lorentz vector $\phi^a$ and singlet $\varphi$ 0-forms. $R={\rm D}\omega$ is the curvature 2-form of the Lorentz connection and D = d+$\omega$ is the Lorentz covariant derivative. 
In the potential term, $\Lambda_4$ is a dimensionless Lagrange multiplier 4-form, $v$ is a dimensionful parameter and $g$ is a self-coupling.
For simplicity we assume $\phi^a$ are dimensionless and $\varphi$ has mass dimension 1. A length scale $l$ is introduced to make $c_1$, $c_2$ $c_3$ dimensionless constants. They will later be absorbed in observables. $\epsilon$ is the totally anti-symmetric invariant of $SO(1,3)$ and little Latin indices run from 0 to 3. Furthermore, the fields transform under local Poincare transformation as follows
\ba\label{trans-omega} \delta\omega^a\!_b &=& {\rm D}\lambda^a\!_b,\\
\label{trans-e}\delta e^a &=& \lambda^a\!_b e^b-{\rm D}\varepsilon^a,\\
\label{trans-phi}\delta\phi^a &=& \lambda^a\!_b\phi^b -\varepsilon^a\varphi,\\
\label{trans-varphi}\delta\varphi &=& 0,\ea
which leave the action intact. $\lambda$ and $\varepsilon$ are parameters of Lorentz transformations and translations respectively. 

It is useful to note that there are 45 (=24+16+4+1) degrees of freedom. There are 10+4 gauge conditions, 4 constraints and 24 on-shell conditions from equations of motion (see below) which removes 42 degrees of freedom. There remain 2+1 decoupled propagating degrees of freedom. Compared to General Relativity, there are 4 St\"{u}ckelberg fields for local translations $\phi^a$ which are total gauge redundancies and the singlet $\varphi$. 
\paragraph*{The Equations of Motion}
The equation of motion of the scalar $\varphi$ reads as follows (group indices and $\epsilon$ are removed for brevity)
\ba\label{vphi-eom} 0 &=& -2g\Lambda_4\varphi(\varphi^2-v^2)\nn\\
&+& c_1 R\wedge R \nn\\ &+& c_2R\wedge({\rm D}\phi+\varphi e)\wedge ({\rm D}\phi+\varphi e) \nn\\ &+& c_3 ({\rm D}\phi+\varphi e)\wedge({\rm D}\phi+\varphi e)\wedge ({\rm D}\phi+\varphi e)\wedge({\rm D}\phi+\varphi e)\nn\\
&+& 2 c_2 \varphi R\wedge ({\rm D}\phi+\varphi e) \wedge e\nn\\
&+& 4 c_3 \varphi ({\rm D}\phi+\varphi e) \wedge ({\rm D}\phi+\varphi e) \wedge ({\rm D}\phi+\varphi e) \wedge e,\nn\\
&-& 2 c_2 \phi \varphi R \wedge e \wedge {\rm d}\varphi\nn\\
&-&12 c_3 \phi\varphi ({\rm D}\phi+\varphi e) \wedge ({\rm D}\phi+\varphi e) \wedge e \wedge {\rm d}\varphi\nn\\
&+& 2c_2 R\wedge (R\phi+\varphi T+{\rm d}\varphi\wedge e)\phi \nn\\
&+& 2c_2 R\wedge ({\rm D}\phi+\varphi e)\wedge {\rm D}\phi\nn\\
&+& 12c_3 ({\rm D}\phi+\varphi e) \wedge ({\rm D}\phi+\varphi e) \wedge (R\phi+\varphi T+{\rm d}\varphi\wedge e)\phi\nn\\
&+& 4c_3 ({\rm D}\phi+\varphi e) \wedge ({\rm D}\phi+\varphi e) \wedge ({\rm D}\phi+\varphi e)\wedge  {\rm D}\phi.\ea
The equation of motion of $\phi^a$ fields is derived as
\ba\label{phi-eom} 0&=& 
-2c_2R\wedge ({\rm D}\phi+\varphi e) \wedge {\rm d}\varphi\nn\\
&-& 4c_3 ({\rm D}\phi+\varphi e)\wedge ({\rm D}\phi+\varphi e)\wedge ({\rm D}\phi+\varphi e)\wedge {\rm d}\varphi\nn\\
&-& 2c_2 \varphi R \wedge (R\phi+\varphi T + {\rm d}\varphi\wedge e)\nn\\
&-& 12c_3 \varphi({\rm D}\phi+\varphi e)\wedge({\rm D}\phi+\varphi e)\wedge (R\phi+\varphi T + {\rm d}\varphi\wedge e)\nn\\
&+& 2 c_2 R\wedge {\rm d}\varphi\wedge {\rm D}\phi\nn\\
&+& 12 c_3 \phi ({\rm D}\phi+\varphi e)\wedge (R\phi+\varphi T + {\rm d}\varphi\wedge e)\wedge {\rm d}\varphi\nn\\
&+& 4 c_3 ({\rm D}\phi+\varphi e)\wedge ({\rm D}\phi+\varphi e) \wedge {\rm d}\varphi\wedge {\rm D}\phi.\ea
The equation of motion of connection $e^a$ is computed as
\ba\label{e-eom} 0&=& 2c_2 \varphi\varphi R\wedge({\rm D}\phi+\varphi e)\nn\\
&+& 4c_3 \varphi\varphi ({\rm D}\phi+\varphi e)\wedge({\rm D}\phi+\varphi e)\wedge({\rm D}\phi+\varphi e)\nn\\
&-& 2c_2 \varphi\phi R\wedge {\rm d}\varphi\nn\\
&-& 12 c_3\varphi\phi ({\rm D}\phi+\varphi e)\wedge ({\rm D}\phi+\varphi e)\wedge {\rm d}\varphi.\ea
Finally, the equation of motion of the  connection $\omega^{ab}$ is
\ba\label{omega-eom} 0&=& 2c_2\varphi\phi R\wedge ({\rm D}\phi+\varphi e)\nn\\
&+& 4c_3\varphi\phi({\rm D}\phi+\varphi e)\wedge ({\rm D}\phi+\varphi e)\wedge ({\rm D}\phi+\varphi e)\nn\\
&-& 2c_2 \phi\phi R\wedge {\rm d}\varphi\nn\\
&-& 12 c_3\phi\phi ({\rm D}\phi+\varphi e)\wedge ({\rm D}\phi+\varphi e)\wedge{\rm d}\varphi\nn\\
&-&c_1{\rm d}\varphi\wedge R \nn\\
&-&c_2{\rm d}\varphi\wedge({\rm D}\phi+\varphi e)\wedge({\rm D}\phi+\varphi e)\nn\\
&-&2c_2\varphi({\rm D}\phi+\varphi e)\wedge(R\phi+\varphi T + {\rm d}\varphi\wedge e)\nn\\
&+& 2c_2({\rm D}\phi+\varphi e)\wedge{\rm d}\varphi\wedge{\rm D}\phi\nn\\
&+& 2c_2 (R\phi+\varphi T + {\rm d}\varphi\wedge e)\wedge {\rm d}\varphi\phi.\ea
\paragraph*{The General Relativity Gauge}
We can use the translational part of the gauge symmetry to choose a gauge in which $\phi^a=0$. Then, the action \eqref{Poincare-action} reduces to the Einstein-Hilbert action plus Euler term plus cosmological constant. Given that $\varphi=v$ one finds $\mpl^2=2lc_2v^3$ and $\Lambda_{cc} = c_3lv^4$. Now $e^a$ can be interpreted as the vierbein and the spacetime metric can be defined as $e^a_\mu e^b_\nu\eta_{ab}$.
In this limit the action is invariant under the residual local Lorentz transformations and diffeomorphisms which are indeed the gauge symmetries of General Relativity. 
\subsection*{Masses from Spontaneous Symmetry Breaking}
Now we study the vacuum solution in which the local symmetries of the action \eqref{Poincare-action}  is spontaneously broken in the following pattern 
\be ISO(1,3)\times {\it Diff} \rightarrow SO(1,3)_D,\ee 
where  $SO(1,3)_D$ is a diagonal subgroup of  and the Lorentz group and an $SO(1,3)\subset GL(4)\subset {\it Diff}$ (Similar symmetry breaking pattern has been studied in \cite{Goon:2014paa,Gabadadze:2013ria}
 in the framework of non-linear realization. However, no dynamical mechanism is presented there). 

We are interested in solutions where $\varphi$ receives a vacuum expectation value by the potential term
\be \langle \varphi\rangle = v.\ee
In this vacuum, the equation of motion of $e^a$ \eqref{e-eom} implies
\ba\label{sol-e} c_2  R + 2c_3 ({\rm D}\phi+\varphi e)\wedge({\rm D}\phi+\varphi e) &=& 0.\ea
Using that, the equation of motion of $\omega^{ab}$ \eqref{omega-eom} gives
\be\label{sol-omega} R^{ab}\phi_b+T^a \varphi = 0.\ee
Given \eqref{sol-e} and \eqref{sol-omega} the equation of motion of $\phi^a$ \eqref{e-eom} is solved with no extra constraint. Finally, the equation of motion of $\varphi$ requires that
\be\label{sol-vphi} 2c_1  R + c_2 ({\rm D}\phi+\varphi e)\wedge({\rm D}\phi+\varphi e)=0.\ee 
It can be seen that \eqref{sol-e} and \eqref{sol-vphi} put a constraint on the action parameters as 
\be \label{constant-relation} 4c_1c_3=c_2^2.\ee 
Eventually, we need to solve a single equation \eqref{sol-e}. It can be seen there is a stable non-trivial solution to \eqref{sol-e} such that
\be\label{the-solution} \langle{\rm D}\phi^a \rangle = u^{(a)}\, {\bf 1}^a,\ee
where ${\bf 1}^a$ is a constant 1-form. For instance on torsion free spacetime $T^a=0$ (thus $\phi^a$ is a covariantly constant vector) we have ${\rm d}e^a = c^{abc}(x)e^b\wedge e^c$ so $\omega^{ab} = g^{abc}(x)e^c$ and thus $R^{ab}=f^{abcd}(x)e^c\wedge e^d$. Then, equation \eqref{sol-e} implies that $u$ has to be a constant and furthermore fixes $(c_3/c_2)=F(u,v)$.

In components we can write \eqref{the-solution} as $\langle D_\mu\phi^a \rangle = u\, \delta_\mu^a$.
This solutions obviously breaks Lorentz invariant and diffeomorphisms. However, the break down of Lorentz symmetry can be compensated by the rotation part of diffeomorphisms  such that a diagonal subgroup survives. 

Now it can be seen that the dRGT potential is obtained in the following form from dynamical symmetry breaking
\ba\label{potential} -V = lc_3v \epsilon_{abcd}(u^{(a)}{\bf 1}^a + ve^a)&\wedge&(u^{(b)}{\bf 1}^b + ve^b) \nn\\ \wedge (u^{(c)}{\bf 1}^c+ ve^c)&\wedge&(u^{(d)}{\bf 1}^d+ ve^d).\ea
This very particular antisymmetric potential removes the ghost like Boulware-Deser mode from the spectrum \cite{Hinterbichler:2012cn}. In this vacuum besides the five polarization modes of massive graviton (three of which are  Goldstone modes from $\phi^a$), there is an extra scalar mode from $\varphi$. 

The Planck constant is defined as follows
\be \mpl^2 = 2 lc_2v^3 .\ee
Furthermore, comparing the dRGT potential \eqref{dRGT-potential} with \eqref{potential} we find that (after defining dimensionless parameters $\alpha_i$ and common $u$ as $\alpha_iu^i\equiv{\rm f}(u^{(a)})$, $i=1,2,3$)
\ba 24\alpha_1lc_3v^4u&=&b_1m^2\mpl^2,\\
24\alpha_2lc_3v^3u^2 &=&b_2m^2\mpl^2,\\
24\alpha_3lc_3v^2u^3 &=&b_3m^2\mpl^2.\ea
Then the graviton mass can be read as follows
\be m^2 = 12 (c_3/c_2)(u/v)(\alpha_3u^2+2\alpha_2uv+\alpha_1v^2),\ee
Consistency checks (avoiding tachyonic graviton and satisfying \eqref{constant-relation}) imply that sign$(c_1,c_2,c_3,v,u)$ is either of 
$(---++), (+++++), (+-++-), (-+--+), (-----), (+++--)$. 
We remind the reader that in the dRGT theory in order that flat space is a solution one gets  $b_0+3b_1+3b_2+b_3=0$. Given that one finds $b_0m^2=12(c_3/c_2)u^2$, this constraint is satisfied for $(v/u)^3+3(v/u)^2\alpha_1+3(v/u)\alpha_2+\alpha_3=0$. 

\subsection*{Action from Contraction}
In this section we show that the action \eqref{Poincare-action} can be obtained from a topological de Sitter gauge theory (for early works on dS gauge theory see \cite{MacDowell:1977jt,Chamseddine:1976bf,Stelle:1979aj,Pagels:1983pq,Grignani:1991nj}).  
The theory is invariant under local de Sitter transformations and spacetime diffeomorphisms  $SO(1,4)\times {\it Diff}$. It is useful to look at this as a principal fibre bundle with $SO(1,4)$-valued bundles over 4 dimensional base (spacetime) manifold on which diffeomorphisms act. 
The action is a functional of the the connection 1-form A$^{AB}$ and a (dimensionless) dynamical field $\phi^A$ in the vector representation. The capital latin indices run from 0 to 4.
The action is
\ba\label{dS-action} S = \epsilon_{ABCDE}\int_{\cal M} &c_1&F^{AB}\wedge F^{CD}\phi^E\nn\\
+ &c_2&F^{AB}\wedge{\it D}\phi^C\wedge {\it D}\phi^{\rm D}\phi^E\nn\\
+ &c_3&{\it D}\phi^A\wedge {\it D}\phi^B\wedge {\it D}\phi^C\wedge {\it D}\phi^{\rm D}\phi^E\nn\\
-\int_{\cal M}&&g(\eta^{AB}\phi_A\phi_B-v^2)^2\Lambda_4.\ea
$F$ is the curvature of the connection
\be F^{AB} = {\it D}{\rm A}^{AB} = {\rm d}{\rm A}^{AB} + {\rm A}^A\!_C\wedge{\rm A}^{CB},\ee
and de Sitter covariant derivative is defined as follows
\be {\it D}\phi^A = {\rm d}\phi^A + {\rm A}^A\!_B\phi^B.\ee
In the above action, $\Lambda_4$ is a dimensionful 4-form Lagrange multiplier, $\eta_{AB}$ is the de Sitter invariant numerical  tensor and $v$ is some dimensionless constant. 

In the action \eqref{dS-action} all possible 4-forms are included. However, each term can be multiplied by singlet 0-form powers of $\phi^A$ that we do not study here.

In order to make connection with the action \eqref{Poincare-action} we formally decompose the dynamical fields into $SO(1,3)$ representations as 
\ba {\rm A}^{AB} &=& 
\left(
\begin{array}{cc}
\omega^{ab} &\ \ e^a\\
-e^b  &\ 0   
\end{array}
\right),\\
 \phi^A &=& \left(\phi^a\ \ \varphi\right)^T.\ea
Accordingly, the de Sitter curvature is decomposed as
\be F^{AB} =
\left(
\begin{array}{cc}
R^{ab} - e^a\wedge e^b&\ \ T^a\\
-T^b  &\ 0   
\end{array}
\right),
\ee
where $T^a$ is defined as $T^a={\rm D}e^a$.
Moreover, the covariant derivative of the vector becomes
\be {\it D}\phi^A = \left({\rm D}\phi^a + \varphi e^a\quad {\rm d}\varphi-e^a\phi_a\right)^T,\ee
where the Lorentz covariant derivative D defined is as before.
Moreover, now the gauge transformations are 
\ba\label{gauge-trans-omega} \delta\omega^a\!_b &=& {\rm D}\lambda^a\!_b + (e^a\epsilon_b-e_b\epsilon^a),\\
\label{gauge-trans-e}\delta e^a &=& - {\rm D}\epsilon^a+\lambda^a\!_b e^b,\\
\label{gauge-trans-phi}\delta\phi^a &=& \lambda^a\!_b\phi^b+\epsilon^a\varphi,\\
\label{gauge-trans-varphi}\delta\varphi &=& - \epsilon^a\phi_a.\ea
Finally, the de Sitter invariant action \eqref{dS-action} is found as 
\ba\label{decomposed-action} S = &c_1&\epsilon_{abcd}\int_{\cal M}\Big[\varphi(R^{ab}-e^a\wedge e^b)\wedge (R^{cd}-e^c\wedge e^d)  \nn\\ 
&&\qquad\qquad\ +2 \,(R^{ab}-e^a\wedge e^b)\wedge T^c \phi^d\Big]\nn\\
+&c_2&\epsilon_{abcd}\int_{\cal M}\Big[(R^{ab}-e^a\wedge e^b)\wedge ({\rm D}\phi^c+\varphi e^c)\wedge\nn\\ 
&&\qquad\qquad\qquad\qquad\qquad\qquad\qquad({\rm D}\phi^d+\varphi e^d)\varphi \nn\\
&&\qquad\qquad+(R^{ab}-e^a\wedge e^b)\wedge({\rm d}\varphi - \phi^e e^e)\wedge \nn\\
&&\qquad\qquad\qquad\qquad\qquad\qquad\qquad ({\rm D}\phi^c+\varphi e^c)\phi^d \nn\\
&&\qquad\qquad+T^a\wedge ({\rm D}\phi^b+\varphi e^b)\wedge({\rm D}\phi^c+\varphi e^c)\phi^d\Big]\nn\\
+&c_3&\epsilon_{abcd}\int_{\cal M}\Big[\varphi({\rm D}\phi^a+\varphi e^a)\wedge({\rm D}\phi^b+\varphi e^b)\wedge\nn\\ &&\qquad\qquad\qquad\qquad\quad ({\rm D}\phi^c+\varphi e^c)\wedge({\rm D}\phi^d+\varphi e^d) \nn\\
&&\qquad\qquad\  +\phi^a({\rm D}\phi^b+\varphi e^b)\wedge({\rm D}\phi^c+\varphi e^c)\wedge \nn\\
&&\qquad\qquad\qquad\qquad\quad({\rm D}\phi^d+\varphi e^d)\wedge({\rm d}\varphi- \phi^e e_e)\Big]\nn\\
&&\quad-\int_{\cal M}g(\varphi^2+\eta^{ab}\phi_a\phi_b-v^2)^2\Lambda_4.\ea
\paragraph*{Group Contraction}
The Minkowski spacetime can be viewed as infinite radius (vanishing cosmological constant) limit of the de Sitter geometry {\it a.k.a.} the Penrose limit. At the the level of algebra, the Poincare algebra can be obtained from the de Sitter algebra by rescaling some of the generators and taking a singular limit by the In\"{o}n\"{u}-Wigner contraction \cite{Inonu:1953sp}. Both algebras have the same number of generators but the structure constants are different. 
To apply the contraction, we introduce a length scale $l$ and rescale the de Sitter boosts $M^{a5}$ as 
\be l^{-1}M^{a5} = P^a ,\ee
where $P^a$ has dimension mass. It can be easily seen that
\be [P^a,P^b] = l^{-2} M^{ab}\rightarrow 0,\ee
when $l\rightarrow\infty$. The rescaling of the generators is equivalent to the rescaling of the connection. The connection 1-form can be expanded as
\be {\rm A} = {\rm A}^{AB}M_{AB} = \omega^{ab}M_{ab} + (l^{-1}e^a) (l P_a),\ee
where the connection $e^a$ is now dimensionless. Similarly, the transformation parameters are rescaled as 
\be \lambda = \lambda^A\!_B M_A\!^B = \lambda^a\!_b M_a\!^b + (l^{-1} \varepsilon^a)(l P_a),\ee
where now $\varepsilon^a$ has dimension length. Furthermore, we rescale the Lorentz scalar as $\varphi\mapsto l\varphi$ such that now it has mass dimension 1. Finally, We send $l\rightarrow\infty$ and while keeping $\varphi,e$ and $\varepsilon$ fixed. In this limit some parts of the gauge transformations \eqref{gauge-trans-omega} and \eqref{gauge-trans-varphi} change as follows
\ba \delta\omega^a\!_b &=& {\rm D}\lambda^a\!_b + l^{-2}(e^a\epsilon_b-e_b\epsilon^a)\rightarrow {\rm D}\lambda^a\!_b,\\
\delta\varphi &=& - l^{-2}\epsilon^a\phi_a\rightarrow 0.\ea
Together with \eqref{gauge-trans-e} and \eqref{gauge-trans-phi} they form gauge transformations under the Poincare group given in \eqref{trans-omega} to \eqref{trans-varphi}.

One would wonder what would happen at the level of the action (and to the dynamics) when $l\rightarrow\infty$. We substitute the rescaled fields into the the decomposed de Sitter invariant action \eqref{decomposed-action}. We also resale the constants $\Lambda_4\mapsto l^{-3}\Lambda_4$ and $v\mapsto lv$ so that they have dimension mass and keep them fix in the limit $l\rightarrow\infty$. The action can be organized in powers of $l$ as $S = l(\cdots)+l^{-2}(\cdots)+l^{-3}(\cdots)$. Only the terms proportional to $l$ survive in $l\rightarrow\infty$ limit and those give exactly the Poincare invariant action \eqref{Poincare-action}.

\subsection*{Conclusion}
In this note we argued that a Lorentz invariant massive gravity (\`{a} la dRGT) can be obtained dynamically from spontaneous symmetry breaking in a topological Poincare gauge theory. In the broken phase, besides polarization modes of massive graviton, there is a Higgs-like scalar excitation which can extend the theory of massive spin 2 into a Lorentz invariant UV completion. The explicit computations of the tree-level  forward scattering of 4 longitudinal modes is studied somewhere else \cite{Torabian:2017}. 

\paragraph*{Acknowledgments}
This work is partially supported by the School of Particles and Accelerators of Institute for Studies in Fundamental Sciences (IPM) and the Research Office of Sharif University of Technology. The author grateful to the International Centre for Theoretical Physics, Trieste for hospitality and to Seif Randjbar-Daemi for fruitful discussions.

\ \\$^*$ Email: {\tt mahdi@physics.sharif.edu}

\end{document}